\documentstyle[aps,twocolumn,floats,epsf]{revtex}

\begin{document}  

\draft

\title{Generation of Intrinsic Vibrational 
Gap Modes in Three--Dimensional
Ionic Crystals}
         
\author{S.~A.~Kiselev$^{*}$ and A.~J.~Sievers} 

\address{Laboratory of Atomic and Solid State Physics and the
Materials Science Center Cornell University, Ithaca, New York
14853-2501}

\maketitle
     
\begin{abstract}
The existence of anharmonic localization of lattice vibrations in a
perfect 3-D diatomic ionic crystal is established for the rigid-ion
model by molecular dynamics simulations.  For a realistic set of NaI
potential parameters, an intrinsic localized gap mode vibrating in the
[111] direction is observed for fcc and zinc blende lattices.  An
axial elastic distortion is an integral feature of this mode which
forms more readily for the zinc blende than for the fcc structure.
Molecular dynamics simulations verify that in each structure this
localized mode may be stable for at least 200 cycles.
\end{abstract}     
\par                                         
\vspace{0.5cm}

%\narrowtext

It has been proposed~\cite{Dolgov86,Sievers88} and numerically
demonstrated~\cite{Burlakov90} that a large amplitude vibration in a
perfect 1-D lattice can localize because of anharmonicity.  More
detailed analytical and numerical investigations of classical 1-D
anharmonic chains made possible by simple eigenvalue generating
recursion relations have revealed a variety of stable intrinsic
localized modes with frequencies outside of the plane wave
bands
\cite{Burlakov90,Page90,Sandusky92,Kivshar92,Kiselev95,Franchini96}.
Recently quantum mechanical aspects of ILMs (Intrinsic Localized
Modes) have been considered~\cite{Rossler95,Wang96}. Some progress
also has been reported for higher dimensional classical crystal
lattices with simple nearest-neighbor model
interactions~\cite{Takeno90,Flach94}.  Particularly important has been
the recognition that diatomic crystal potentials like the
Born--Mayer--Coulomb in 1-D produces an intrinsic gap mode (IGM) {\em
between} the optic and acoustic branches instead of an ILM above the
plane wave spectrum~\cite{Kiselev93}.  The possibility of IGMs in 3-D
anharmonic lattices with realistic potentials has remained elusive.
One approach has been to focus on crystal surfaces and edges where
harmonic localization already plays an important
role~\cite{Teixeira95,Bonart95a,Bonart95b}.

In this Letter we demonstrate with molecular dynamics simulations
that, for sufficiently large vibrational amplitude, anharmonicity can
stabilize an IGM in a 3-D uniform diatomic crystal with rigid ion NaI
potential arranged in either the fcc or zinc blende structure.  By
developing a self-consistent numerical technique for finding an
intrinsic localized mode eigenvector, we have been able to show that
for a given gap mode amplitude with the same potential that the
localization is much stronger for a $T_d$ symmetry site when compared
to an $O_h$ one.

To construct the stationary localized mode eigenvector for the
nonlinear 3-D diatomic lattice with long range interactions, we build
on techniques which have been used successfully to identify 1-D
anharmonic modes~\cite{Kiselev93}, one of which is the rotating wave
approximation.  A Fourier extension of that idea for vibration in a
stationary periodic mode with fundamental frequency, $\omega$, which
includes the static and second harmonic for the $i^{th}$ particle
displacement, ${\bf r}_i(t)$, is

\begin{equation}
{\bf r}_i(t) = \sum_{n=0}^{2} {\bf r}_i^{(n)} \cos n \omega t \;,
\label{eqn:1}
\end{equation}
where ${\bf r}_i^{(0)}$, is the distorted equilibrium position of the
$i^{th}$ particle, and the ${\bf r}_i^{(n)}$'s are the time-independent
amplitudes of the different harmonics. The force acting on the
$i^{th}$ particle can also be represented by a similar Fourier series.
Substituting the coordinate and force Fourier series into the
classical equations of motion and equating the terms with different
harmonics gives a time-independent system of nonlinear equations

\begin{equation}
{\bf F}_i^{(n)} + n^2 \omega^2 m_i {\bf r}_i^{(n)} = 0 \;,
\;\;\;\; n = 0, 1, 2 \;.
\label{eqn:2}
\end{equation}
The Fourier coefficients, ${\bf F}_i^{(n)}$, are determined in the
usual way~\cite{Kiselev93}.

The procedure for finding the anharmonic localized mode eigenvector
relies on first generating a fictitious dynamics for the Fourier
amplitudes and then applying a version of dynamical simulated
annealing~\cite{Car85}.  The coefficients ${\bf \tilde{r}}_i^{(0)}$,
${\bf \tilde{r}}_i^{(2)}$, and ${\bf \tilde{r}}_i^{(1)}$ 
are now taken to
vary with time and obey the following fictitious equations of motion:

\begin{eqnarray}
{\bf \tilde{F}}_i^{(n)} & = & 
{\bf F}_i^{(n)} + n^2 \omega^2 m_i {\bf \tilde{r}}_i^{(n)} =
m_i \frac{d^2 {\bf \tilde{r}}_i^{(n)}}{dt^2} \;,
\\ \nonumber
n & = & 0, 1, 2 \;;
\label{eqn:3}
\end{eqnarray}
so that dynamical simulated annealing can be used to find the
equilibrium values for these quantities ${\bf r}_i^{(0)}$, ${\bf
r}_i^{(2)}$, and ${\bf r}_i^{(1)}$, when all 
${\bf \tilde{F}}_i^{(n)}\left({\bf r}_i^{(n)}\right)=0$.  
Our iterative procedure to obtain the eigenvector is
as follows:

\begin{enumerate}

\item
\label{item:1}
Guess an initial shape for the gap mode eigenvector.  In our case we
use the mass-defect eigenvector associated with a harmonic gap mode of
frequency $\omega$.  The rescaled mass-defect gap mode eigenvector
gives initial amplitudes ${\bf \tilde{r}}_i^{(1)}$ where  
$r_0^{(0)}=\alpha$ while 
${\bf \tilde{r}}_i^{(0)}$, ${\bf \tilde{r}}_i^{(2)}$, 
and the initial velocities are set to zero.

\item
\label{item:2}
Make an MD time step by solving Eqs.~(3) and update
the quantities, ${\bf \tilde{r}}_i^{(n)}$ and the ${\bf
\dot{\tilde{r}}}_i^{(n)}$ for fixed central particle amplitude
$\alpha$.  The classical equations of motion are integrated using the
``leap--frog'' algorithm~\cite{Hockney81} with a time step of 1.35 fs.

\item
\label{item:3}
Apply a form of dynamical simulated annealing~\cite{Car85}.  During a MD
simulation run, the oscillatory values of the kinetic energies of each
of these three objects ${\bf \tilde{r}}_i^{(n)}$ per particle for the
$N$ particles is monitored.  (Each object vibrates around its
equilibrium position ${\bf r}_i^{(n)}$).  When the total kinetic
energy of the system of objects passes through its maximal value in
the MD fictitious time-evolution, all object's velocities 
${\bf \dot{\tilde{r}}}_i^{(n)}$ are set to zero.  In this way we
incrementally move the system closer to its equilibrium configuration.

\item
\label{item:4}
After 40 MD steps as described in~\ref{item:2} and checking for
condition~\ref{item:3} the intrinsic gap mode frequency $\omega$ is
updated by solving the single eq. 
$F_0^{(1)}+\omega^2m_0r_0^{(1)}=0$ for $\omega$ with $r_0^{(1)}=\alpha$.

\item
\label{item:5}
Verify the correctness and stability of the resultant IGM eigenvector
with a regular MD simulation.  Repeat~\ref{item:1}
through~\ref{item:4} until the lifetime of the mode remains unchanged
when the resultant eigenvector is used as an initial condition for an
MD simulation.  As long as the frequency of the mode remains in the
gap we find the procedure described here converges to a localized
eigenvector.  No change in the results is observed when this procedure
is repeated with a 1/2 time step.

\end{enumerate}

To investigate the possibility of a 3-D intrinsic gap mode in ionic
crystals we use the tabulated rigid ion potential for NaI from Table's
2 and 4 of Ref.~\cite{Michielsen75}.  
The potential has the following form

\begin{eqnarray}
\phi\left(r_{ij}\right) & = &
\frac{z_i z_j e^2}{4 \pi \epsilon_0 r_{ij}} +
\frac{b}{r_{ij}^4} \exp \left[ -k\left( r_{ij}-r_{0ij}\right)\right] 
\\ \nonumber
& & - \frac{c_{ij}}{r_{ij}^6} -
\frac{d_{ij}}{r_{ij}^8}
\label{eqn:interact-potential}
\end{eqnarray}
with $r_{ij}$ the distance between the ions $i$ and $j$, $z_i$ and
$z_j$ are $\pm 1$ charges, $\epsilon_0$ is the permittivity of vacuum,
$r_{0ij}$ is the sum of the ionic radii, $b$ and $k$ are parameters
determined by fitting the thermal expansion and isothermal
compressibility, $c_{ij}$ and $d_{ij}$ are respectively the
coefficients for the dipole--dipole and dipole--quadrupole
interactions.  The lattice constant for the fcc structure is chosen to
be $a=6.35\AA$, within 1\% of the experimental value [20] .  The
calculated TO frequency is $\omega_{TO}=2.51\times 10^{13}$~rad/s,
within 5\% of the experimental value of 
$2.39\times 10^{13}$~rad/s~\cite{Sangster78}.
The gap between the optic and acoustic branches extends from
$\omega_{+}=2.41\times 10^{13}$~rad/s to $\omega_{-}=1.46\times
10^{13}$~rad/s.

In carrying out the MD calculations on a 216-ion cube with periodic
boundary conditions the method of Sangster and Dixon~\cite{Sangster76}
has been used to evaluate the Ewald sum.  For this method the cut-off
distance in real space is approximately half the length, $L/2$, of the
cube (the interaction of sixth neighboring shell is counted), and the
reciprocal lattice is summed with the convergence parameter of
$5.6/L$.

The resulting intrinsic gap mode eigenvector for the fcc NaI crystal
is shown in Fig.~1.  Panel (a) identifies the first harmonic part of
the vibrational eigenvector, ${\bf r}_i^{(1)}$, localized on a central
light ${\rm Na^+}$ ion and its neighboring shells with the central
${\rm Na^+}$ ion vibrating in the [111] direction.  The vibrational
amplitude of the central ${\rm Na^+}$ ion to the nn distance, $d$,
$r_0^{(1)}/d=\alpha/d=0.244$.  The gap mode frequency lies 5.0\% below
the bottom of the optic band.  The elastic distortion is characterized
by the dc part of the mode's eigenvector, ${\bf r}_i^{(0)}$, which is
shown in panel (b) of Fig.~1

\begin{figure}

\vspace{-0.3in}
\begin{center}
\leavevmode
\epsfxsize=9cm
\epsfbox{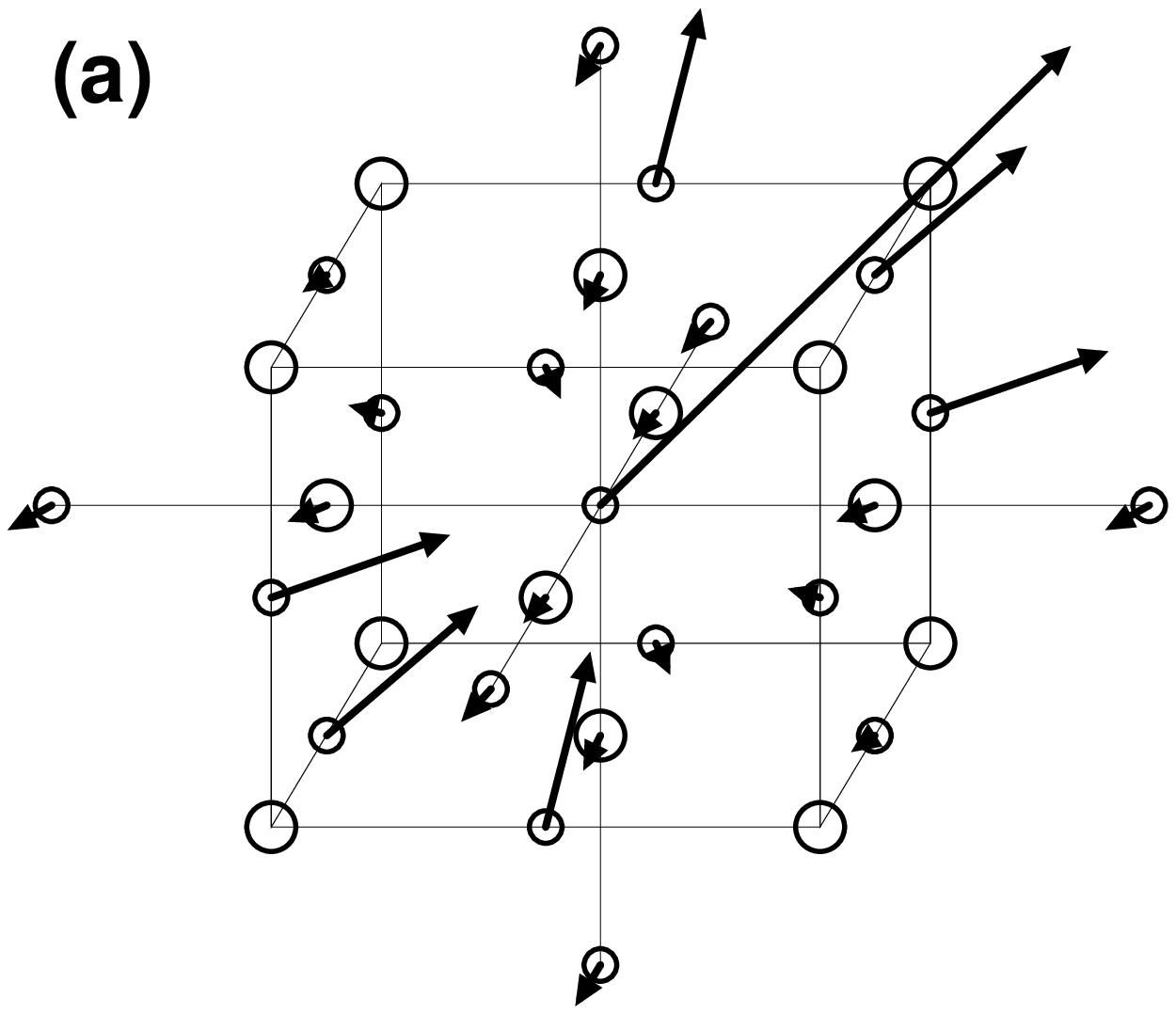} 
\end{center}

%\vspace{-0.5in}
%\begin{center}
%\leavevmode
%\epsfxsize=9.0cm
%\epsfbox{fig1.eps}
%\end{center} 

\vspace{-0.7in}
\begin{center}
\leavevmode
\epsfxsize=9.0cm
\epsfbox{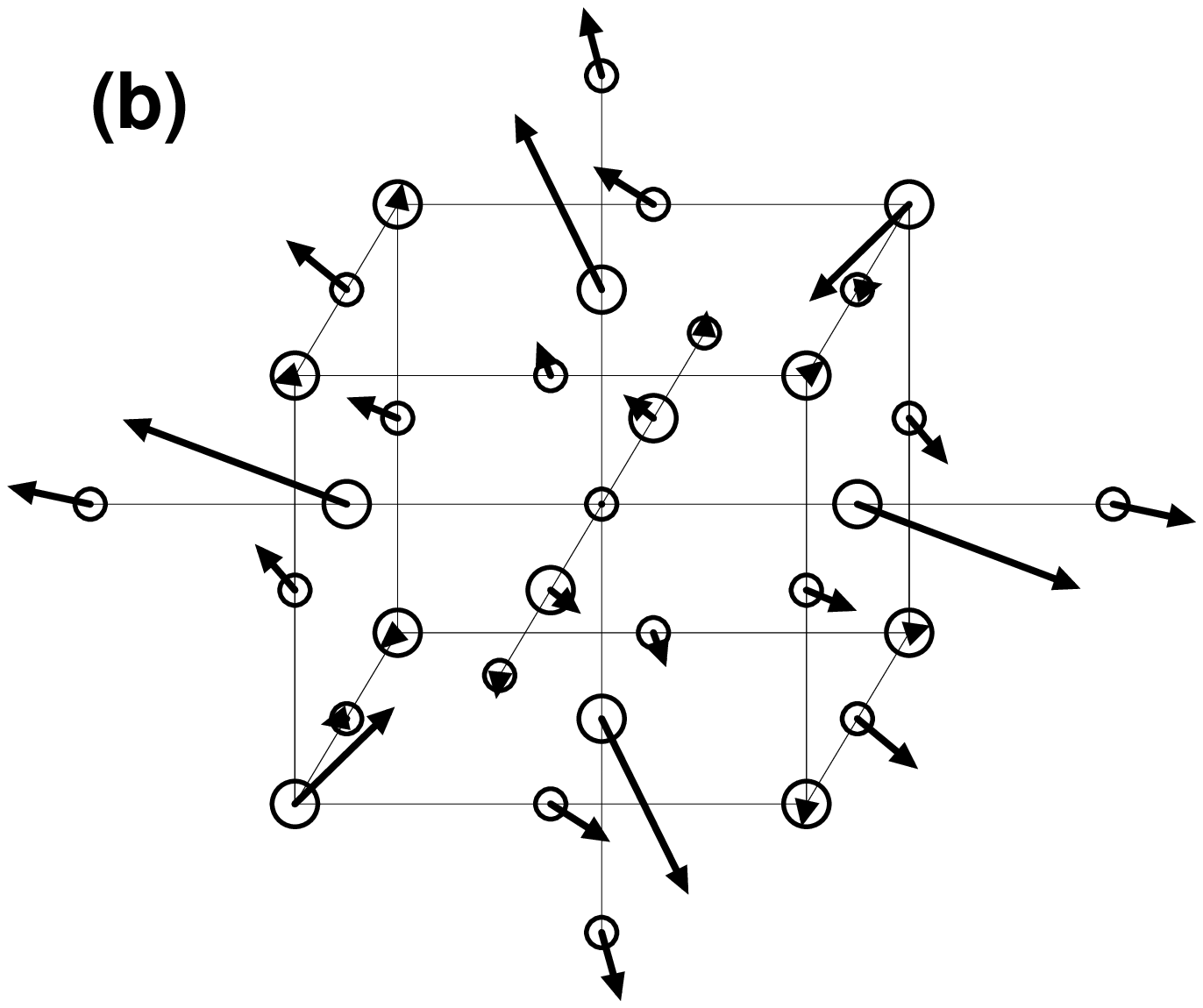}
\end{center} 

\vspace{-0.3in}
\caption{The intrinsic gap mode eigenvector 
in the NaI crystal with fcc structure.  The mode has a relative
amplitude $\alpha/d=0.244$ and a relative frequency
$\omega/\omega_+=0.950$.  Panel (a) shows the first harmonic part of
the eigenvector.  The arrows represent 10~$\times$ actual amplitudes
for clarity.  Panel (b) shows the dc part of the eigenvector.  The
arrows representing the displacements are expanded 40~$\times$ for
clarity.  Small circles, ${\rm Na^+}$ ions; large circles, ${\rm I^-}$
ions.  Although not clear from the figure the magnitudes of the dc
distortion for the nearest ${\rm I^-}$ ion shell are the same,
consistent with an axial distortion around the [111] axis.}

\label{fig:1}
\end{figure}

The above procedure has also been performed on a larger array of
particles to insure that the periodic boundary conditions do not
influence the results.  In a 1000 ion crystal the IGM with amplitude
$\alpha/d=0.244$ has the same eigenvector as shown in Fig.~1 and its
frequency differs by 1\% from that for the 216 ions NaI crystal.  This
frequency difference is associated with a slightly different crystal
distortion between the two cases.  The MD simulation test shows that
the IGM remains stable in the large crystal and its life-time
increases slightly.

\begin{figure}

\vspace{-0.3in}
\begin{center}
\leavevmode
\epsfxsize=9cm
\epsfbox{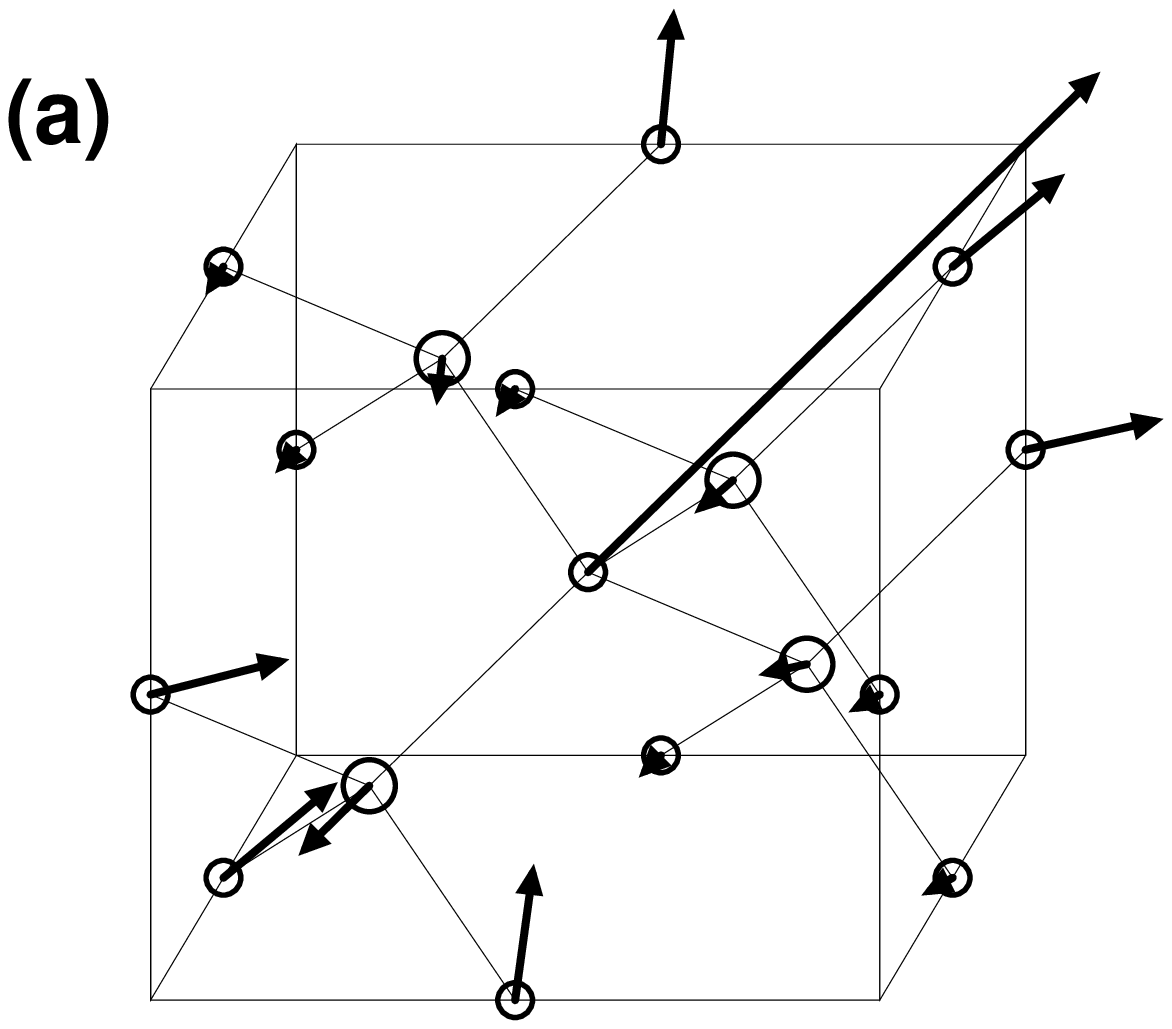} 
\end{center}

%\vspace{-0.5in}
%\begin{center}
%\leavevmode
%\epsfxsize=9.0cm
%\epsfbox{fig2.eps}
%\end{center} 

\vspace{-0.7in}
\begin{center}
\leavevmode
\epsfxsize=9.0cm
\epsfbox{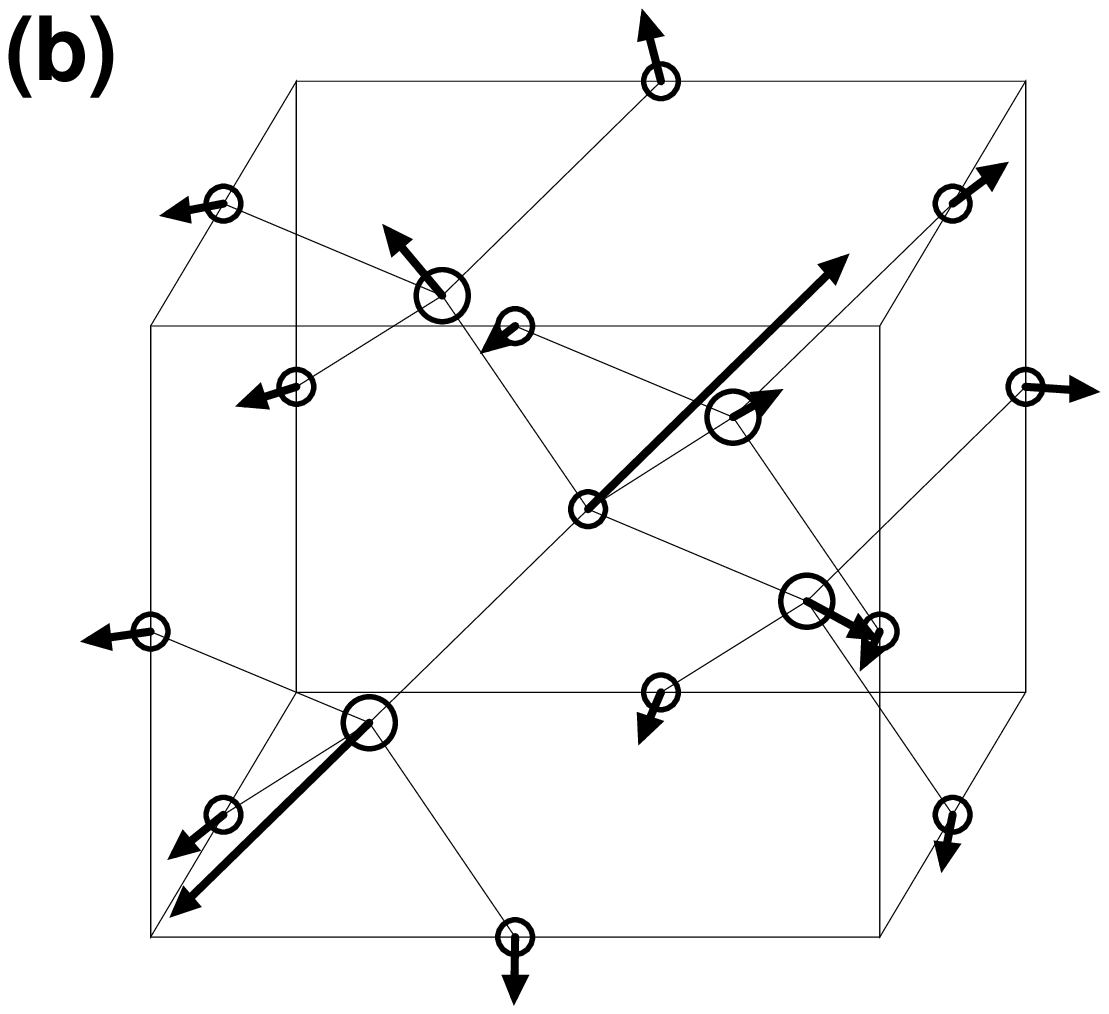}
\end{center} 

\vspace{-0.3in}
\caption{The intrinsic gap mode eigenvector 
for a hypothetical NaI crystal with zinc blende structure.  The mode
has a relative amplitude $\alpha/d=0.116$ and a relative frequency
$\omega/\omega_+=0.950$.  Panel (a) shows the first harmonic part of
the eigenvector.  The arrows represent 20~$\times$ actual amplitudes.
Panel (b) shows the dc part of the eigenvector.  The arrows
representing the displacements are expanded 40~$\times$.  Small
circles, ${\rm Na^+}$ ions; large circles, ${\rm I^-}$ ions.}

\label{fig:2}
\end{figure}

In order to investigate the role of point group symmetry on the
intrinsic gap mode parameters the zinc blende structure has also been
tested with the same potential.  (A local minimum of the lattice
energy occurs at the slightly larger lattice constant, $a=7.00\AA$.)
The IGM eigenvector for 216 particles is shown in Fig.~2.  Again the
central ion vibrates in the [111] direction.  This mode has an
amplitude to nn distance, $r_0^{(1)}/d=\alpha/d=0.116$.  Again the
mode vibrational eigenvector, ${\bf r}_i^{(1)}$, is localized on a
central light ${\rm Na^+}$ ion and its neighboring shells.  Note that
although the relative mode amplitude is only one half that of the IGM
shown in Fig.~1, its relative frequency occurs at the same value in
the gap (5.0\%) due to the larger elastic distortion in the zinc
blende lattice near the mode center. The maximum vibrational amplitude
and dc distortion for each of the shells of particles shown in Figs.~1
and 2 are given in Table~\ref{table:1}.

\begin{table}[h]

\caption{Maximum amplitude in a shell versus shell index 
for Figures 1 and 2. Both the vibrational amplitude and the 
dc distortion are given as fractions of the nn distance.}

\begin{tabular}{lllll}
                & fcc &  & zinc blende &
\\ 
  shell        & vib. ampl. & dc dist. & vib. ampl. & dc dist.
\\ \tableline 
0 & 0.2438	& 0		& 0.1156	& 0.0291 \\
1 & 0.0119	& 0.0244	& 0.0147	& 0.0220 \\
2 & 0.0734	& 0.0074	& 0.0242	& 0.0064 \\
3 & 0.0025	& 0.0132	& 0.0005	& 0.0039 \\
4 & 0.0149	& 0.0078	& 0.0039	& 0.0038 
\end{tabular}

\label{table:1}
\end{table}

Typical power spectra of the central particle vibration is presented
in Fig.~3 for the eigenvectors shown in Figs.~1 and 2 after about 200
vibrations.  These spectra reflect the stable vibration of the IGM at
frequencies close to the values predicted by
Eq.~(\ref{eqn:2}) both for the fcc and zinc blende
structures.  Note that the vibrational mode for the $O_h$ symmetry site
shows a weak third harmonic while the IGM for the $T_d$ symmetry site
shows all harmonics.

\begin{figure}

\vspace{-0.3in}
\begin{center}
\leavevmode
\epsfxsize=9cm
\epsfbox{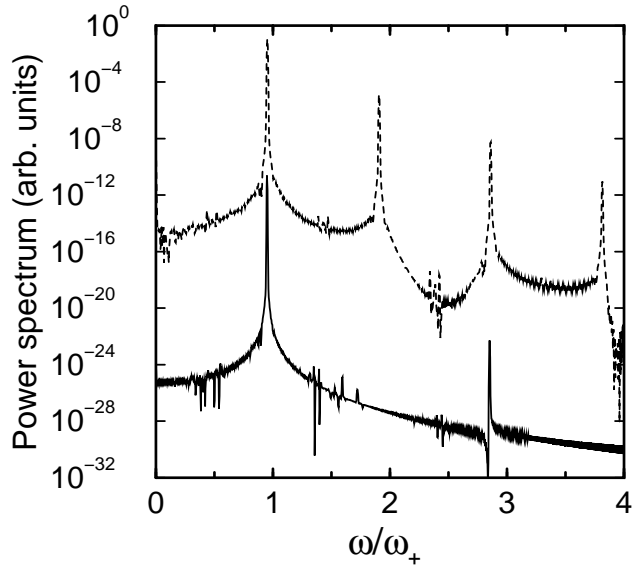} 
\end{center}

\vspace{-0.1in}
\caption{The power spectrum of the central particle vibration 
for an IGM in an NaI crystal with either fcc or zinc blende structure.
The eigenvectors shown in Figs.~1 and 2 are used as the initial
conditions for the MD simulations.  The solid curve (fcc lattice) is
shifted down by 10 decades from the dashed curve (zinc blende) for
clarity.  The noise is associated with truncational errors.}

\label{fig:3}
\end{figure}

Figure~4 shows the IGM frequency verses the amplitude for the two
structures under investigation.  The left sets of data are for zinc
blende and the right sets are for the fcc lattice.  These results
indicate that the $T_d$ symmetry site appears to support more
anharmonicity in the sense that for a given vibrational amplitude the
frequency of the IGM drops farther into the forbidden gap and has a
larger elastic lattice distortion around the IGM center [compare
Figs.~1(b) and 2(b)].

\begin{figure}

\vspace{-0.3in}
\begin{center}
\leavevmode
\epsfxsize=9cm
\epsfbox{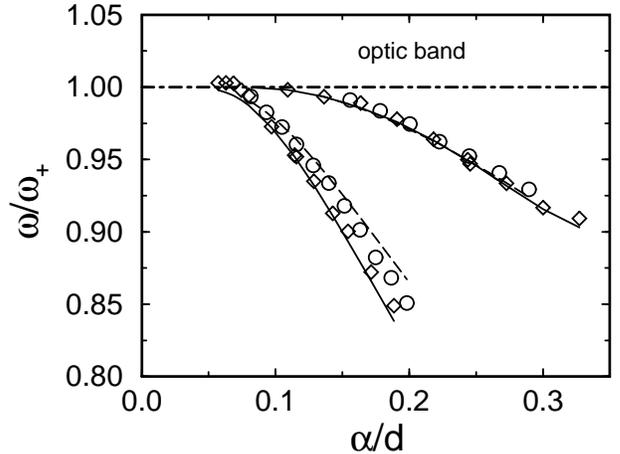} 
\end{center}

\vspace{-0.3in}
\caption{The frequency of an intrinsic gap mode 
as a function of normalized amplitude, $\alpha/d$.  The left set of
data are for the zinc blende and the right set for the fcc lattice.
The numerical solutions of Eq.~(\ref{eqn:2}) are
represented by the solid lines for the [111] and by dashed lines for
the [110] crystal direction.  The result of MD simulations are given
by the open diamonds for the [111] direction and by open circles for
the [110] direction.}

\label{fig:4}
\end{figure}

The general behavior of the IGM frequency versus amplitude shown in
Fig. 4 is similar for both structures --- with increasing mode
amplitude the frequency drops farther into the gap.  An amplitude
threshold is evident.  An IGM which has its frequency about 5\% below
the bottom of the optics band (corresponding to the middle regions of
the curves in Fig.~4) has the longest lifetime of $\sim$200--250
vibrational periods.  If its frequency drops farther into the
forbidden gap ($\sim$10\%) the mode's lifetime decreases to
$\sim$100 periods.  At the opposite small amplitude limit its
life-time again decreases ($\sim$40 periods) presumably because of the
strong coupling between the IGM and the nearby plane waves.
  
In both systems the elastic distortion associated with the IGM has
lower symmetry than the corresponding point group symmetry of the
crystal.  To recover the full crystal point group symmetry for this
perfect crystal it must be possible with a different set of initial
conditions to rotate the IGM about the equilibrium lattice site.  To
test this idea we have examined IGM excitations along the three
different crystal directions.  Although a stable IGM mode does not
appear for initial excitation along the [100] direction, the dashed
lines and open circles in Fig.~4 demonstrate that for both point group
symmetries a [110] directed mode can occur.  These sets of data
support the idea of an interchange of the IGM vibration direction as
would be expected, for example, with hindered rotational motion of the
excitation about the lattice site.  This concomitant low frequency
component of the IGM may provide new experimental ways to excite and
identify these nonlinear excitations.

\par                                         
\vspace{0.5cm}

We thank S.~R.~Bickham, G.~V.~Chester, R.~H.~Silsbee and M.~P.~Teter
for helpful discussions.  This work is supported in part by
NSF-DMR-931238, ARO-DAAH04-96-1-0029 and the MRL central facilities.
Some of this research was conducted using the resources of the Cornell
Theory Center, which receives major funding from the National Science
Foundation and New York State, with additional support from other
members of the center's Corporate Partnership Program.

\end{document}